\documentclass[preprint,12pt]{elsarticle}
\usepackage{graphics}
\usepackage{lscape}
\usepackage{hyperref}
\usepackage{xfrac}
\usepackage{graphicx} 
\usepackage[font=small,labelfont=bf]{caption} 
\usepackage[subrefformat=parens]{subcaption}
\usepackage{amssymb}
\usepackage{multirow}
\usepackage[LGRgreek]{mathastext}

\begin{document}

\begin{frontmatter}

\title{Investigation of long lived activity produced due to neutron emitting reactions}

 \author[label1,label2]{Tanmoy Bar}
 \author[label1,label2]{Dipali Basak}
 \author[label1,label2]{Sukhendu Saha}\author[label1,label2]{Lalit Kumar Sahoo}
 \author[label1]{Chinmay Basu}
 \address[label1]{Saha Institute of Nuclear Physics, 1/AF, Bidhannagar, Kolkata-700064,INDIA}
 \address[label2]{Homi Bhabha National Institute, Anushaktinagar, Mumbai-400094, INDIA}

\begin{abstract}
In this article, a detailed investigation has been done for the long lived gamma activity due to neutron emitting experiments. These calculations mainly focused on the experiments used for energy calibration purposes. Calibrated energy is one of the most essential features of any accelerator facility. Several experiments have been used for this purpose. Generally, experiments having sharp curvature change in cross section of yield are used. Neutron emitting experiments are one of such. Around the globe reactions like $^7Li(p,n)$, $^{13}C(p,n)$, $^{19}F(p,n)$, $^{27}Al(p,n)$ etc. are used to calibrate energy of the beam with accelerator terminal voltage. Neutrons coming from these experiments can interact with surrounding elements. These interactions with neutrons can create long lived gamma activity which may interfere with future measurements. The present study has been done keeping in mind the new Facility for Research in Experimental Nuclear Astrophysics (FRENA) at Saha Institute of Nuclear Physics. It is a 3MV tandetron low energy high current machine.

\end{abstract}
\end{frontmatter}

\section{Introduction}
Energy calibration is the first and most essential step in a newly installed accelerator facility. There are several experiments performed by different accelerator facilities across the globe for energy calibration such as $^{7}Li(p,n)$, $^{13}C(p,n)$, $^{27}Al(p,n)$, $^{27}Al(p,\gamma)$, $^{19}F(p,n)$ etc.. Among which neutron emitting reactions are very popular due to their relatively easier detections. Every neutron producing reaction has a particular energy, called neutron threshold energy which is the minimum required energy of the incident beam to produce neutron. The present study has been done keeping in mind the Facility for Research in Experimental Nuclear Astrophysics (FRENA)\cite{Ref1} located at Saha Institute of Nuclear Physics, INDIA. FRENA is a 3 MV Tandetron low energy high current accelerator dedicated for low energy Nuclear Astrophysics research. This is a new machine and energy calibration is needed before doing any experiment. Few neutron-producing reactions are generally chosen for calibration purposes. Neutrons generated from such experiments can interact with any elements present in the accelerator hall and produce different isotopes. Now, if produced isotopes are radioactive in nature and have a long half-life (for few months to several years) then those isotopes will contribute as background gamma peaks in the next performing experiments. These background peaks will overlap with the gammas from the actual experiments if the energy of those background gammas is very close to the gammas from actual astrophysical experiments. A systematic study has been done for all the possible radioactive isotopes that may produce during some experiments which are planned to perform in FRENA for energy calibration of the machine.

\section{Energy calibration for FRENA accelerator}
The terminal voltage of FRENA can be varied from 200 kV to 3 MV.  The energy of the charged particle produced in the ion source is accelerated by this terminal voltage. 
The energy of any charged particle in a potential difference is given by the relation:

\begin{equation}
    E_{particle} = (q+1) \times V.   
\end{equation}

Where, $E_{particle}$ is the charged particle energy, q is the charge state of the projectile, V is the terminal voltage.

In practical cases, the accelerated charge passes through several magnetic fields and finally bombard into a target material. Since every machine has some intrinsic deviation from the ideal value. Energy needs to be calibrated so that a more accurate relation between terminal voltage and beam energy can be established for this particular machine.
\\

\section{Results for different calibration reactions} 
In FRENA, available proton energy varies from 400 keV to 6 MeV. For energy calibration purposes, those (p,n) experiments (Table. 1) are chosen for which neutron threshold energy lies within the available energy at FRENA.

\begin{table}
\begin{center}
\begin{tabular}{|p{4cm}|p{2cm}|p{2cm}|p{2cm}|p{2cm}|}
\hline

 & $^{7}Li(p,n)$ &$^{13}C(p,n)$ &$^{19}F(p,n)$&  $^{27}Al(p,n)$ \\ \hline
Neutron threshold energy (MeV) & 1.88036 & 3.2355 & 4.23513 & 5.80362 \\ \hline
 
\end{tabular}
\caption{Neutron threshold energy for different (p,n) reactions.}
\end{center}
\end{table} 

 Most of the components of the accelerators are made of 304 and 316 Stainless steel alloy along with copper (Cu) and Tantalum (Ta). Apart from that Zirconium (Zr), Aluminium (Al), Calcium (Ca), Silicon (Si) and Oxygen (O) are also present in the vicinity. Chemical composition for 304 and 316 stainless steel alloys along with maximum \% are listed in Table. 2  and Table. 3.  
\begin{table}
\begin{center}
\begin{tabular}{|p{1cm}|p{0.8cm}|p{0.8cm}|p{0.8cm}|p{0.8cm}|p{0.8cm}|p{0.8cm}|p{0.8cm}|p{1.2cm}|p{0.8cm}|}
\hline
 & C & Mn & Si & P & S & Cr & Ni & Fe & N \\ \hline
Max\% &  0.07 & 2.00 & 1.00 & 0.05 & 0.03 & 19.50 & 10.50 & Balance & 0.11\\ \hline
 
\end{tabular}
\caption{Maximum \% of chemical compositions present in 304 Stainless steel alloy \cite{Ref2}. }
\end{center}
\end{table} 
 
\begin{table}
\begin{center}
\begin{tabular}{|p{1cm}|p{0.5cm}|p{0.8cm}|p{0.8cm}|p{0.8cm}|p{0.8cm}|p{0.8cm}|p{0.8cm}|p{0.8cm}|p{1.2cm}|p{0.8cm}|}
\hline
 & C & Mn & Si & P & S & Cr & Mo &Ni & Fe & N \\ \hline
Max\% &  0.08 & 2.00 & 0.75 & 0.045 & 0.03 & 18.00 &3.00& 14.00& Balance & 0.10\\ \hline
 
\end{tabular}
\caption{Maximum \% of chemical compositions present in 316 Stainless steel alloy \cite{Ref3}. }
\end{center}
\end{table} 

Reaction cross-section of isotopes present in the vicinity with produced neutrons from calibration experiments are calculated using a Hauser-Feshbach statistical model code TALYS 1.95 \cite{Ref4}. In this study, channels having cross-sections less than $10^{-7}$ mb are not considered. Gamma decay schemes of every produced isotope are taken from National Nuclear Data Center (NNDC)\cite{Ref5}. Radioactive isotopes produced by neutrons having a half-life of around 3 months or more are considered. Gammas having a relative intensity of more than 1\% from those radioactive isotopes are listed in this study.

\subsection{$^{7}Li(p,n)$ reaction:}
$^{7}Li(p,n)$ is one of the most common experiments done for energy calibration at accelerator facilities. Proton beam of energy 1.88036 MeV was considered for this study (Table. 1). From kinematics neutron energy for that particular proton beam energy has been calculated and it was found to be between 20 - 42 keV. In calculations, only maximum and minimum values of neutron energy are considered. For the interaction of produced neutron, all the stable isotopes of the elements present in the vicinity are considered. \\
Table. 4  also shows isotopic abundance in nature, reaction product, and its formation cross-section along with gamma energy and corresponding relative intensities are also listed.

 \begin{figure}
  \centering
    \includegraphics[width=.8\linewidth]{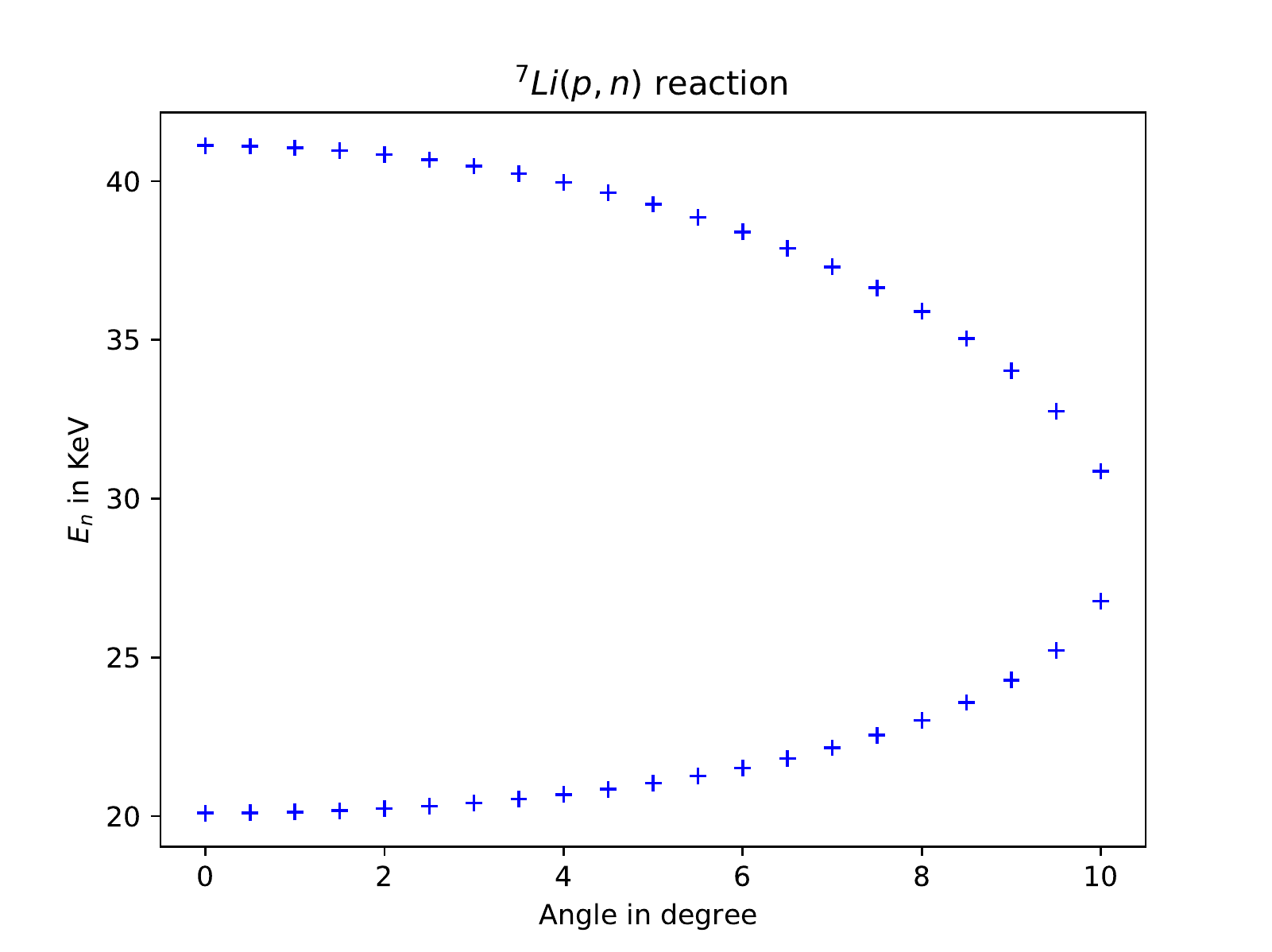}
    \caption{Distribution of neutron energy at threshold energy, $E_p$=1.88036 MeV}
  \end{figure}

\subsection{$^{13}C(p,n)$ reaction:}
A proton beam of energy 3.2355 MeV was considered for the study of this reaction (Table. 1). From kinematics neutron energy for that particular proton beam energy has been calculated and it was found to be between 11 - 24 keV. 
 \begin{figure}
  \centering
    \includegraphics[width=.8\linewidth]{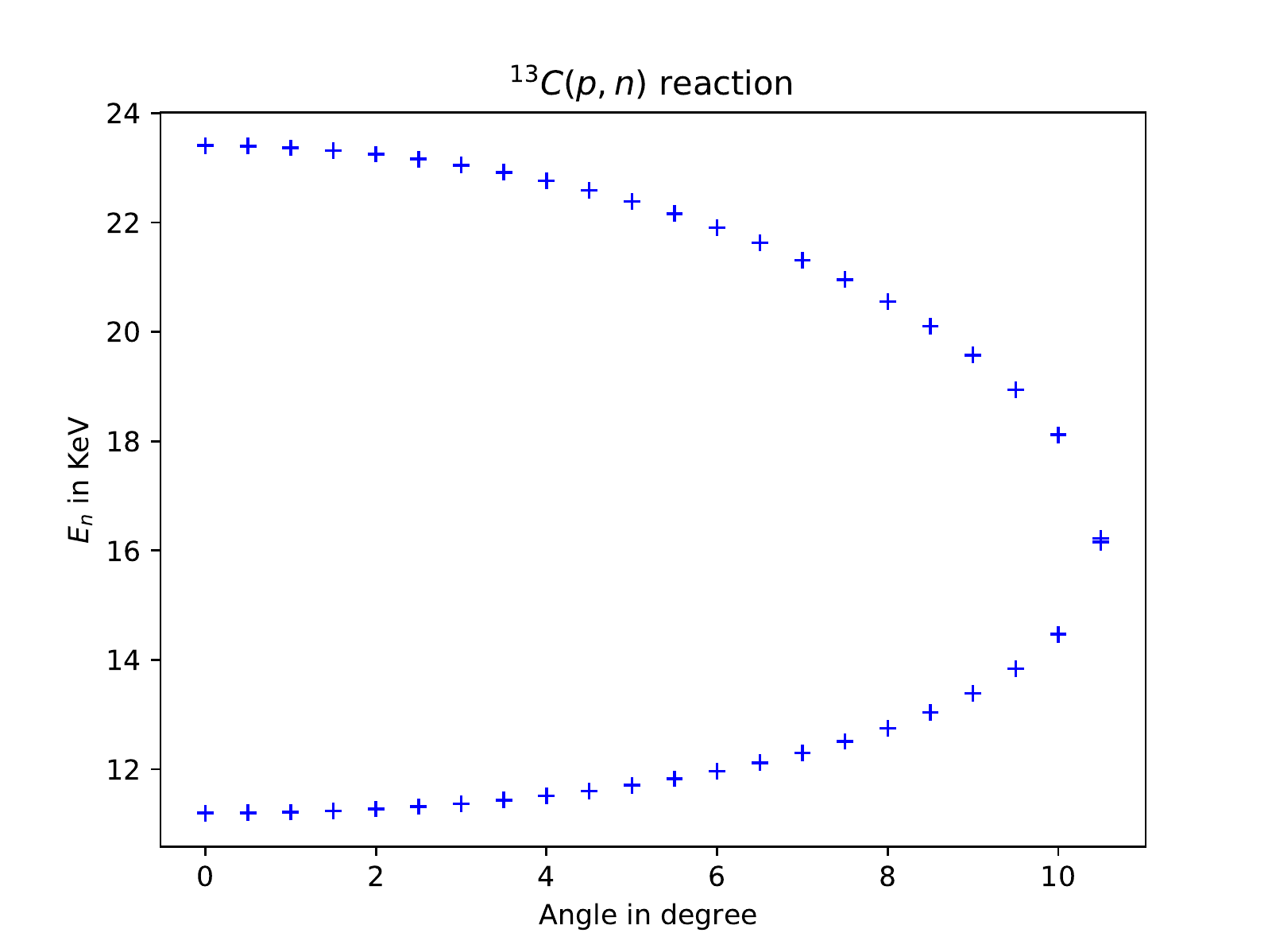}
    \caption{Distribution of neutron energy at threshold energy, $E_p$=3.2355 MeV}
  \end{figure}

\subsection{$^{19}F(p,n)$ reaction:}
Neutron threshold energy for this reaction is 4.23513 MeV. Proton of that energy is considered for this calculation. Neutrons emitted from this reaction ranges between 4 to 20 keV.
 \begin{figure}
  \centering
    \includegraphics[width=.8\linewidth]{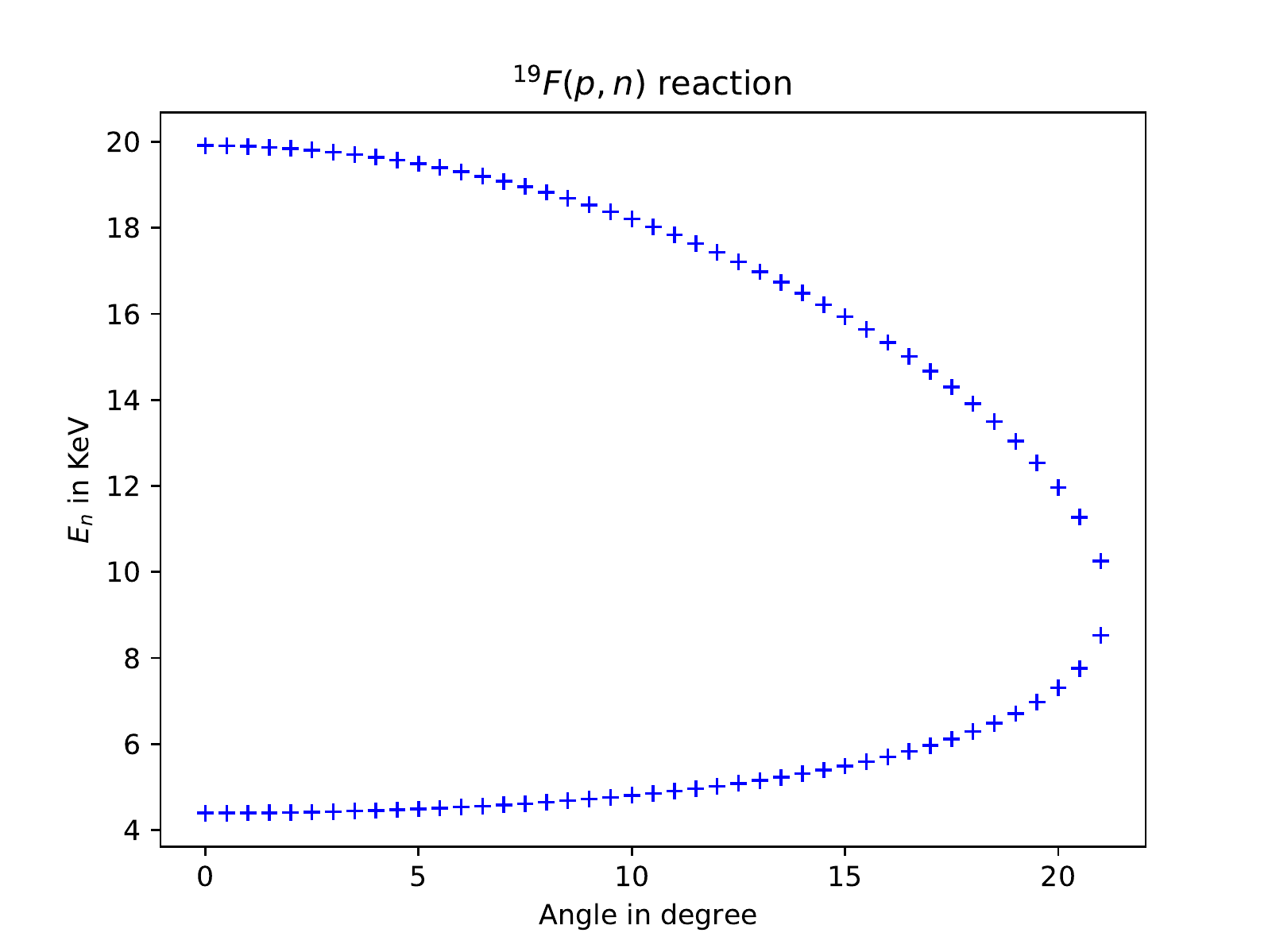}
    \caption{Distribution of neutron energy at threshold energy, $E_p$=4.23513 MeV}
  \end{figure}

\subsection{$^{27}Al(p,n)$ reaction:}

A proton of energy 5.80362 MeV was considered for this study (Table. 1).  From kinematics available  neutron energy for that particular proton beam energy is between 0 - 30 keV. 

 \begin{figure}[h]
  \centering
    \includegraphics[width=.8\linewidth]{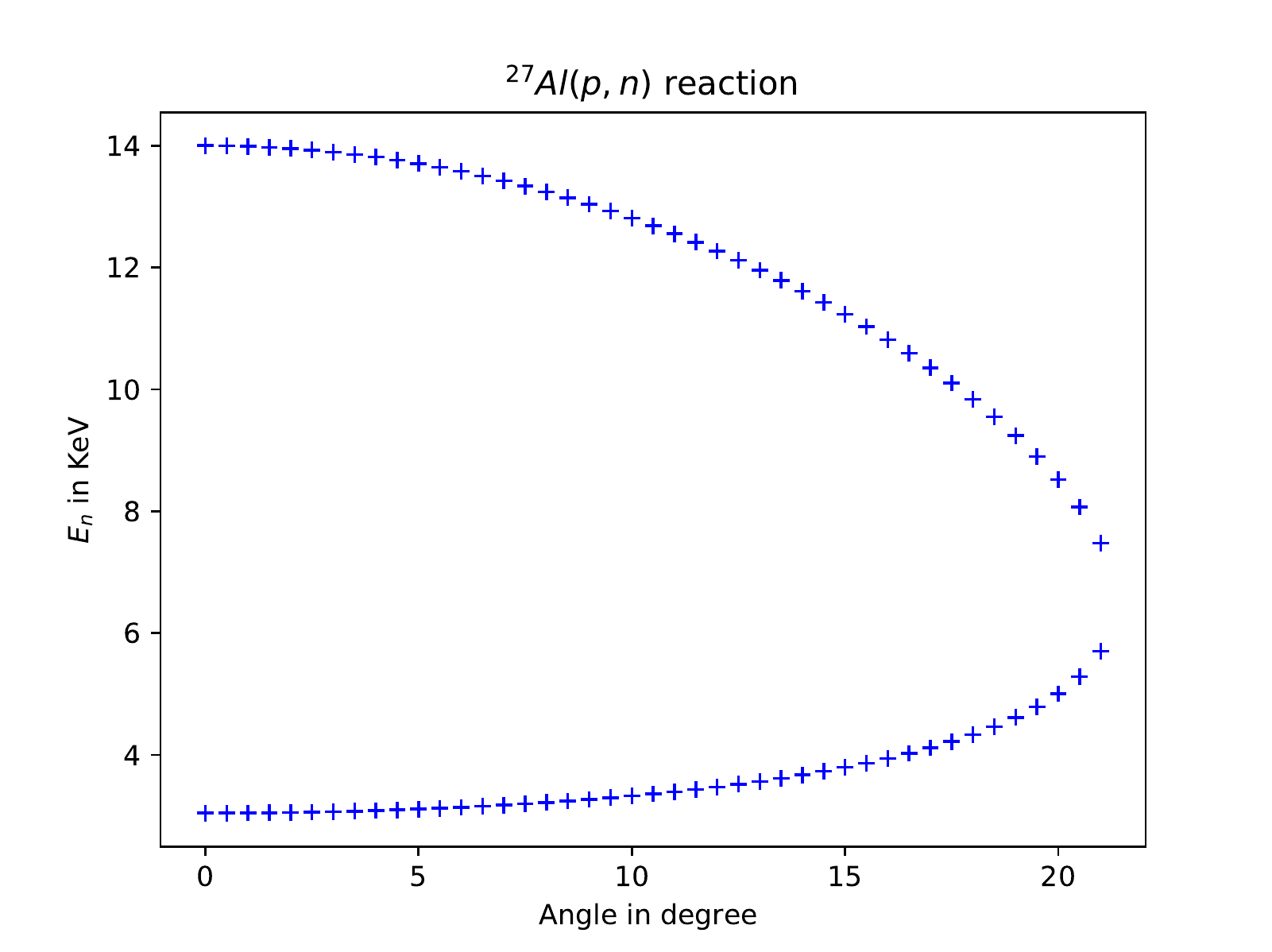}
    \caption{Distribution of neutron energy at threshold energy, $E_p$=5.80362 MeV}
  \end{figure}

Kinematics calculations show that neutron energy for all reactions mentioned above is within the 10 to 40 keV range. Cross-sections of different stable isotopes present in the vicinity of the accelerator hall with the produced neutrons from the calibration reactions are listed in Table. 4. Half-lives of reaction products, daughter nucleus with corresponding gamma energies having relative intensities greater or equal to 1 are also listed in the table.

\begin{table}
\tiny
\begin{center}
\begin{tabular}{|p{.5cm}|p{.5cm}|p{.8cm}|p{.8cm}|p{1cm}|p{1cm}|p{.8cm}|p{1cm}|p{1.2cm}|p{1.5cm}|}
\hline
 Iso-topes & Abun-dance (\%) & Neutron energy (keV)& Reaction product & Cross section (mb)  & half life & Daughter nucleus & Gamma energy (keV) & Gamma intensity (\%) & Remarks\\ \hline \hline

\multirow{2}{*}{$^{13}C$} & \multirow{2}{*}{1.07} & 10  & \multirow{2}{*}{$^{14}C$} &\multirow{1}{*}{1.59$\times 10^{-1}$} & \multirow{2}{*}{5700 y}  & \multirow{2}{*}{$^{14}N$} & & & \multirow{2}{*}{100\% $\beta$ -decay} \\ \cline{3-3} \cline{5-5}
 
 &&20 && 8.61$\times 10^{-2}$ &&&&& \\ \cline{3-3} \cline{5-5}
 &&30 && 5.44$\times 10^{-2}$ &&&&& \\ \cline{3-3} \cline{5-5}
 &&40 && 4.38$\times 10^{-2}$ &&&&& \\ \cline{3-3} \cline{5-5}
\hline
\hline
\multirow{2}{*}{$^{14}N$} & \multirow{2}{*}{99.6} & 10  & \multirow{2}{*}{$^{14}C$} &\multirow{1}{*}{1.28$\times 10^{3}$} & \multirow{2}{*}{5700 y}  & \multirow{2}{*}{$^{14}N$} & & & \multirow{2}{*}{100\% $\beta$ -decay} \\ \cline{3-3} \cline{5-5}
 
 &&20 && 8.68$\times 10^{2}$ &&&&& \\ \cline{3-3} \cline{5-5}
 &&30 && 5.83$\times 10^{2}$ &&&&& \\ \cline{3-3} \cline{5-5}
 &&40 && 5.34$\times 10^{2}$ &&&&& \\ \cline{3-3} \cline{5-5}
\hline
\hline
\multirow{2}{*}{$^{17}O$} & \multirow{2}{*}{0.04} & 10  & \multirow{2}{*}{$^{14}C$} &\multirow{1}{*}{2.80$\times 10^{2}$} & \multirow{2}{*}{5700 y}  & \multirow{2}{*}{$^{14}N$} & & &\multirow{2}{*}{100\% $\beta$ -decay} \\ \cline{3-3} \cline{5-5}
 &&20 && 2.97$\times 10^{2}$ &&&&& \\ \cline{3-3} \cline{5-5}
 &&30 && 1.30$\times 10^{2}$ &&&&& \\ \cline{3-3} \cline{5-5}
 &&40 && 1.89$\times 10^{2}$ &&&&& \\ \cline{3-3} \cline{5-5}
\hline
\hline

\multirow{2}{*}{$^{34}S$} & \multirow{2}{*}{4.25} & 10  & \multirow{2}{*}{$^{35}S$} &\multirow{1}{*}{2.00} & \multirow{2}{*}{87.37 d}  & \multirow{2}{*}{$^{35}Cl$} & & & \multirow{2}{*}{100\% $\beta$ -decay} \\ \cline{3-3} \cline{5-5}
 
 &&20 && 1.216 &&&&& \\ \cline{3-3} \cline{5-5}
 &&30 && 9.41$\times 10^{-1}$ &&&&& \\ \cline{3-3} \cline{5-5}
 &&40 && 7.82$\times 10^{-1}$ &&&&& \\ \cline{3-3} \cline{5-5}
\hline
\hline
\multirow{2}{*}{$^{40}Ca$} & \multirow{2}{*}{96.941} & 10  & \multirow{2}{*}{$^{41}Ca$} &\multirow{1}{*}{5.25$\times 10^{1}$} & \multirow{2}{*}{99400 y}  & \multirow{2}{*}{$^{41}K$} & & & \multirow{2}{*}{100\% $\beta$ -decay} \\ \cline{3-3} \cline{5-5}
 &&20 && 2.46$\times 10^{1}$ &&&&& \\ \cline{3-3} \cline{5-5}
 &&30 && 2.69$\times 10^{1}$ &&&&& \\ \cline{3-3} \cline{5-5}
 &&40 && 1.68$\times 10^{1}$ &&&&& \\ \cline{3-3} \cline{5-5}
\hline
\hline

\multirow{2}{*}{$^{44}Ca$} & \multirow{2}{*}{2.086} & 10  & \multirow{2}{*}{$^{45}Ca$} &\multirow{1}{*}{6.50$\times 10^{2}$} & \multirow{2}{*}{162.61 d}  & \multirow{2}{*}{$^{45}Sc$} & \multirow{2}{*}{12.47}& \multirow{2}{*}{3.0$\times 10^{-6}$}&  \\ \cline{3-3} \cline{5-5}
 &&20 && 4.15$\times 10^{1}$ &&&&& \\ \cline{3-3} \cline{5-5}
 &&30 && 3.25$\times 10^{1}$ &&&&& \\ \cline{3-3} \cline{5-5}
 &&40 && 2.74$\times 10^{1}$ &&&&& \\ \cline{3-3} \cline{5-5}
\hline
\hline

\multirow{2}{*}{$^{54}Fe$} & \multirow{2}{*}{5.85} & 10  & \multirow{2}{*}{$^{55}Fe$} &\multirow{1}{*}{9.46$\times 10^{1}$} & \multirow{2}{*}{2.744 y}  & \multirow{2}{*}{$^{55}Mn$} & \multirow{2}{*}{126}&\multirow{2}{*}{1.28 $\times 10^{-7}$} & \\ \cline{3-3} \cline{5-5}
 
 &&20 && 6.35$\times 10^{1}$ &&&&& \\ \cline{3-3} \cline{5-5}
 &&30 && 5.17$\times 10^{1}$ &&&&& \\ \cline{3-3} \cline{5-5}
 &&40 && 4.56$\times 10^{1}$ &&&&& \\ \cline{3-3} \cline{5-5}
 
\hline
\hline 
\multirow{2}{*}{$^{58}Ni$} & \multirow{2}{*}{68.07} & 10  & \multirow{2}{*}{$^{59}Ni$} &\multirow{1}{*}{2.16$\times 10^{2}$} & \multirow{2}{*}{7.6$\times 10^{4}$ y}  & \multirow{2}{*}{$^{59}Co$} & \multirow{2}{*}{511}&\multirow{2}{*}{7.4 $\times 10^{-5}$} & Annihilation gammas \\ \cline{3-3} \cline{5-5}
 
 &&20 && 1.47$\times 10^{2}$ &&&&& \\ \cline{3-3} \cline{5-5}
 &&30 && 1.20$\times 10^{2}$ &&&&& \\ \cline{3-3} \cline{5-5}
 &&40 && 1.06$\times 10^{2}$ &&&&& \\ \cline{3-3} \cline{5-5}
 
\hline
\hline
\multirow{2}{*}{$^{62}Ni$} & \multirow{2}{*}{3.635} & 10  & \multirow{2}{*}{$^{63}Ni$} &\multirow{1}{*}{5.45$\times 10^{1}$} & \multirow{2}{*}{101.2 y}  & \multirow{2}{*}{$^{63}Cu$} & & & \multirow{2}{*}{100\% $\beta$ -decay} \\ \cline{3-3} \cline{5-5}

 &&20 && 3.78$\times 10^{1}$ &&&&& \\ \cline{3-3} \cline{5-5}
 &&30 && 3.18$\times 10^{1}$ &&&&& \\ \cline{3-3} \cline{5-5}
 &&40 && 2.89$\times 10^{1}$ &&&&& \\ \cline{3-3} \cline{5-5}
\hline
\hline
\multirow{2}{*}{$^{92}Mo$} & \multirow{2}{*}{14.65} & 10  & \multirow{2}{*}{$^{93}Mo$} &\multirow{1}{*}{1.17$\times 10^{2}$} & \multirow{2}{*}{4.0$\times 10^{3}$y}  & \multirow{2}{*}{$^{93}Nb$} & \multirow{2}{*}{30.77}&\multirow{2}{*}{5.20 $\times 10^{-4}$} & \\ \cline{3-3} \cline{5-5}
 
 &&20 && 6.80$\times 10^{1}$ &&&&& \\ \cline{3-3} \cline{5-5}
 &&30 && 5.50$\times 10^{1}$ &&&&& \\ \cline{3-3} \cline{5-5}
 &&40 && 4.05$\times 10^{1}$ &&&&& \\ \cline{3-3} \cline{5-5} 
\hline
\hline 

\multirow{2}{*}{$^{92}Zr$} & \multirow{2}{*}{17.15} & 10  & \multirow{2}{*}{$^{93}Zr$} &\multirow{1}{*}{8.36$\times 10^{1}$} & \multirow{2}{*}{1.61$\times 10^{6}$y}  & \multirow{2}{*}{$^{93}Nb$} & \multirow{2}{*}{30.77}&\multirow{2}{*}{4.30 $\times 10^{-4}$} & \\ \cline{3-3} \cline{5-5}
 
 &&20 && 4.77$\times 10^{1}$ &&&&& \\ \cline{3-3} \cline{5-5}
 &&30 && 3.49$\times 10^{1}$ &&&&& \\ \cline{3-3} \cline{5-5}
 &&40 && 2.83$\times 10^{1}$ &&&&& \\ \cline{3-3} \cline{5-5}
\hline
\hline

\multirow{20}{*}{$^{181}Ta$} & \multirow{20}{*}{99.988} & 10  & \multirow{20}{*}{$^{182}Ta$} &\multirow{1}{*}{2.09$\times 10^{3}$} & \multirow{20}{*}{114.74 d}  & \multirow{20}{*}{$^{182}W$} & \multirow{1}{*}{65.722}&\multirow{1}{*}{3.01} & \\ \cline{3-3} \cline{5-5} \cline{8-9}
 
 &&20 && 1.34$\times 10^{3}$ &&&&& \\ \cline{3-3} \cline{5-5}
 &&30 && 1.05$\times 10^{3}$ &&&&& \\ \cline{3-3} \cline{5-5}
 &&40 && 8.69$\times 10^{2}$ &&&&& \\ \cline{3-3} \cline{5-5}
 
 &&&&&&& 67.749 & 42.9& \\ \cline{8-9}
 &&&&&&& 84.680 & 2.654& \\ \cline{8-9}
 &&&&&&& 100.106 & 14.2& \\ \cline{8-9}
 &&&&&&& 113.672 & 1.871& \\ \cline{8-9}
 &&&&&&& 152.429 & 7.02& \\ \cline{8-9}
 &&&&&&& 156.386 & 2.671& \\ \cline{8-9}
 &&&&&&& 179.393 & 3.119& \\ \cline{8-9}
 &&&&&&& 198.352 & 1.465& \\ \cline{8-9}
 &&&&&&& 222.109 & 7.57& \\ \cline{8-9}
 &&&&&&& 229.321 & 3.644& \\ \cline{8-9}
 &&&&&&& 264.074 & 3.612& \\ \cline{8-9}
 &&&&&&& 1001.7 & 2.086& \\ \cline{8-9}
 &&&&&&& 1121.29 & 35.24& \\ \cline{8-9}
 &&&&&&& 1189.04 & 16.49& \\ \cline{8-9}
 
 &&&&&&& 1221.395 & 27.23& \\ \cline{8-9}
 &&&&&&& 1231.004 & 11.62& \\ \cline{8-9}
 &&&&&&& 1257.407 & 1.51& \\ \cline{8-9}
 &&&&&&& 1289.145 & 1.37& \\ \cline{8-9}
\hline
\hline

\end{tabular}
\caption{Cross-section of different isotopes present in the vicinity of the accelerator with neutrons of different energy, half-lives of produced isotopes and gamma energy with relative intensity
. }
\end{center}
\end{table}

\clearpage
\section{Discussion and conclusion}
In this work, a systematic detailed study has been done to identify any possible long-lived gamma emitting isotopes formed during the calibration study. Apart from the above listed isotopes two more isotopes $^{100}Mo$  and $^{180m}Ta$ (0.077 MeV, 9-) are produced in the vicinity during the energy calibration. These two isotopes have no decay record in NNDC. Decay of $^{180m}Ta$ having half-life  $>4.5 \times 10^{16} y$ has never been observed till date\cite{Ref6} \cite {Ref7}.  $^{100}Mo$ having half-life $6.8 \times 10^{18} y$ undergoes two-neutrino double $\beta$ -decay \citep{Ref8}. Reaction products such as $^{55}Fe$, $^{59}Ni$, $^{45}Ca$, $^{93}Mo$, $^{93}Zr$ have half-lives much greater than 3 months but intensities of gammas from those nuclei have very low intensity. Isotopes like $^{63}Ni$, $^{14}C$, $^{35}S$, $^{41}Ca $ have significantly long half-lives. These isotopes undergo 
$\beta-$ decay, with no record of any gamma decay channel in NNDC. This study shows that experiments mentioned above for the energy calibration purpose are producing only $^{182}Ta$ which has significant long-lived activity (114.72 days) and decay gamma over a wide range of energies (60 - 1300 keV) with an intensity of more than 1\%. several gamma energy over a wide range. Gamma of energies 1121.29, 1189.04, 1221.395, 1231.004 keV having intensities 35.24, 16.49, 11.23, 11.62 respectively can have a significant effect on experimental data if performed just after the calibration experiments. Nuclear astrophysics experiments detecting gamma after the calibration study needs to be spaced enough to minimize the effect of such background gammas from long-lived isotopes especially if the expected gamma energy of the performing experiments overlap with the background energy.
  \\

\section*{Acknowledgement}

\end{document}